\documentclass[12pt]{elsarticle} 

\usepackage[a4paper, margin=3cm]{geometry} 
\usepackage{setspace} 
\usepackage{amsmath,amssymb,amsfonts}
\usepackage{algorithm}
\usepackage{algpseudocode}
\usepackage{graphicx}
\usepackage{textcomp}
\usepackage{xcolor}
\usepackage{caption}
\usepackage{tabularx}
\usepackage{array}
\usepackage{comment}
\usepackage{float}
\usepackage{subfig}
\usepackage{booktabs}
\usepackage{array}
\usepackage{enumitem}
\usepackage{hyperref} 
\usepackage{makecell}
\usepackage[font=normalsize]{caption}
\begin{document}

\begin{frontmatter}
\title{A Review of 
Community-Centric Power System Resilience: Strategies, Data-Driven Methods, and Techno-Legal Perspectives}

\author[aff1]{Masoud H. Nazari\corref{cor1}}
\author[aff1]{Hamid Varmazyari}
\author[aff1]{Antar Kumar Biswas}

\author[aff2,aff3]{Ümit Cali}

\author[aff4]{Hollis Belnap}
\author[aff4]{Masood Parvania}

\cortext[cor1]{Corresponding author: Masoud H. Nazari (email: masoud.nazari@wayne.edu)}

\address[aff1]{Department of Electrical and Computer Engineering, 
Wayne State University, Detroit, Michigan, USA}

\address[aff3]{Department of Physics, Engineering and Technology, University of York, York, United Kingdom}

\address[aff2]{Artificial Intelligence, Smart Infrastructure and Robotics Research Center, American University of Sharjah, Sharjah, United Arab Emirates}

\address[aff4]{Department of Electrical and Computer Engineering,
University of Utah, Salt Lake City, Utah, USA}

\begin{abstract}

This paper presents a comprehensive review of community-centric power system resilience, emphasizing the integration of community-level resilience considerations and techno-legal governance frameworks with engineering-based resilience enhancement strategies and data-driven approaches to address extreme events. Recent large-scale outages have demonstrated that power disruptions can cascade beyond electrical infrastructure and disproportionately affect vulnerable communities, critical services, and interconnected urban systems, highlighting the need for resilience approaches that integrate technical, social, and regulatory dimensions. Within this community-centric perspective, the review first summarize state-of-the-art strategies for enhancing power system resilience, including network hardening, resource allocation, optimal scheduling, and system reconfiguration techniques, while highlighting the growing role of artificial intelligence (AI) and data-driven analytics in supporting resilience planning and operational decision-making. It then examines the interdependencies between power system resilience and community resilience, addressing socioeconomic and behavioral dimensions, cross-infrastructure interconnections, and the emerging role of resilience hubs. The paper further examines the techno-legal frameworks governing resilient energy systems by comparing the regulatory landscapes of the European Union (EU) and the United States, highlighting key similarities and distinctions that shape resilience planning and implementation. 
By analyzing state-of-the-art engineering-based, AI-driven, and techno-legal methods for assessing and mitigating the impacts of high-impact, low-probability (HILP) events, the review identifies critical research gaps 
and outlines promising directions for future investigation.
\end{abstract}


\end{frontmatter}

\section{Introduction}
\begin{figure}[t!]
    \centering
    \includegraphics[width=.7\textwidth]{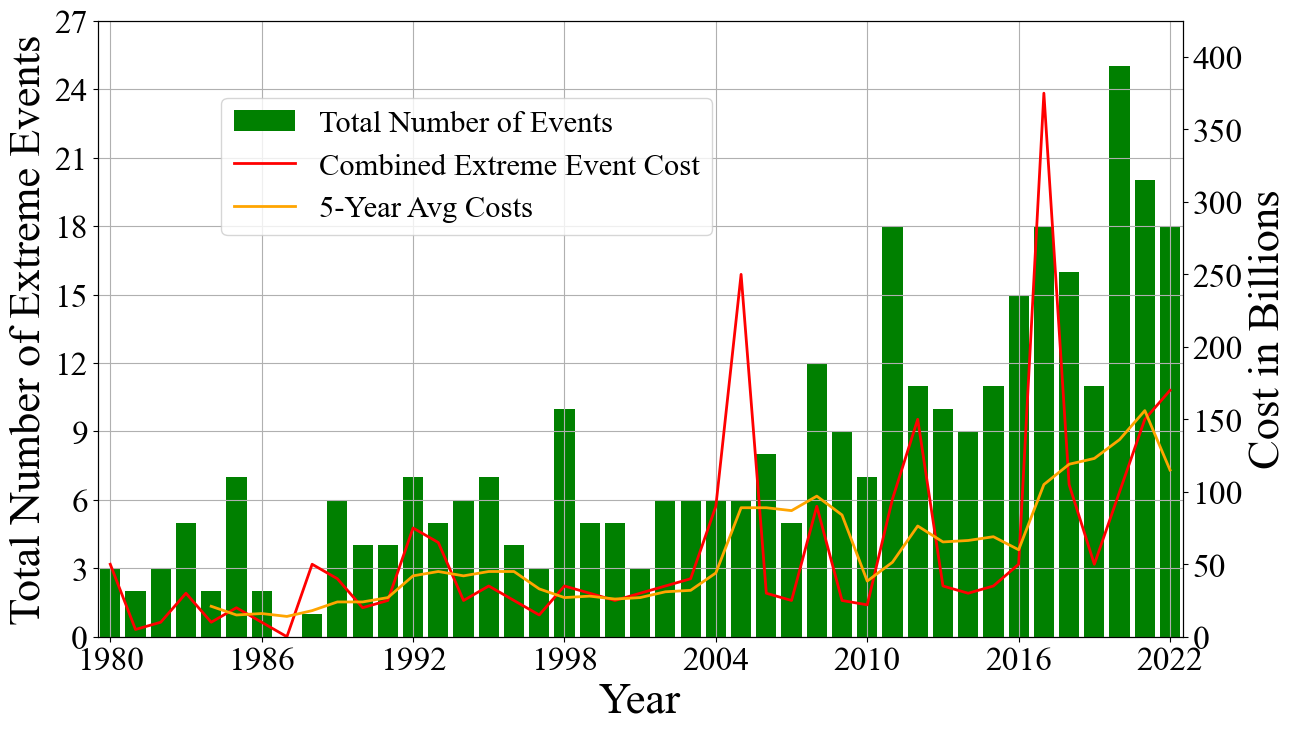}
    \caption{Frequency and associated economic loss of U.S. extreme weather events.}
    \label{economic loss}
    \vspace{-1em}
\end{figure}

The frequency of power outages and their associated losses have been growing in the U.S. over the past 43 years, as illustrated in Fig.~\ref{economic loss}~\cite{climategov}. 
Aging infrastructure, increasing system complexity, and the rising intensity of extreme events have exposed critical vulnerabilities in modern power systems~\cite{mukherjee2018multi}. 
Storm-related disruptions are estimated to cost the U.S. economy between \$20~billion and \$55~billion annually~\cite{campbell2012weather}. Historical events further highlight the societal consequences of prolonged outages.
For example, Hurricane Maria in 2017 caused a widespread collapse of the electric
power system in Puerto Rico, leaving many communities without electricity for
months and contributing to an estimated 2,975 excess deaths 
\cite{kishore2018maria_mortality}.
Similarly, the 2021 Texas winter storm (Storm Uri) triggered widespread power
outages affecting more than 4.5 million customers 
disproportionately impacting low-income households and
critical community services \cite{Texas21}. More 
recent large-scale outages, including the April~2025 Iberian Peninsula blackout and the 2025 California blackout, demonstrate how extreme-event-driven power failures can cascade beyond the electric grid, disrupting transportation and communications systems ~\cite{sf_blackout_waymo_2025,entsoe_blackout_2025}. 
Several studies further anticipate that both the frequency and severity of such high-impact events will continue to increase under climate change~\cite{campbell2012weather,panteli2016boosting,sadeghi2019warning}. 
This underscores the critical need to adapt power grids to withstand extreme disturbances and mitigate catastrophic impacts.

In this context, power system resilience has emerged as a key framework for guiding the analysis and design of modern power grids. It refers to the ability of the system to withstand, adapt to, and rapidly recover from disruptive events, while also encompassing broader socio-economic dimensions through the concept of community resilience~\cite{ribeiro2015enhancing}. 

Power system resilience assessment can generally be categorized into two complementary paradigms: component-level engineering approaches and system-level data-driven approaches. Engineering-based methods rely on fragility modeling of individual components—such as transmission lines, substations, and generation assets—to quantify failure probabilities under extreme events and evaluate physical vulnerability scenarios~\cite{panteli2016power,ranjbar2021resiliency}. 
While these approaches provide valuable insight into failure mechanisms, their practical application is often constrained by limited access to detailed component-level data. In contrast, data-driven approaches evaluate resilience using aggregate performance metrics—such as resilience curves—to characterize system-level degradation and recovery~\cite{bruneau2003framework,panteli2017metrics,jufri2019state,biswas5151316data}.

Prior studies have shown that community resilience is shaped by multiple socioeconomic and demographic factors, which also influence power system resilience and recovery outcomes~\cite{loni2024data,dominianni2018power}. 
Moreover, power systems are tightly interdependent with other critical infrastructures, including transportation, communication, and healthcare systems, such that failures in one domain can propagate and amplify impacts across others. 
These interdependencies highlight the importance of coordinated resilience planning to maintain stability in integrated energy systems~\cite{zhang2016modeling,bagchi2010modeling}. Similar ideas are increasingly discussed in the smart infrastructure and smart city literature, where resilience is framed as a socio-technical challenge \cite{adeleke2025smart}.

A substantial body of literature has reviewed resilience concepts, assessment frameworks, and enhancement strategies for power systems \cite{ahmadi2021frameworks,jufri2019state,bhusal2020power,mishra2021review}. However, most existing reviews primarily focus on technical resilience enhancement
methods or analytical assessment approaches, while comparatively limited attention
has been given to integrating these strategies with community vulnerability
considerations and the techno-legal frameworks that govern resilience planning and implementation. As power grids become increasingly digitized and interconnected with other critical infrastructures, resilience planning must address not only engineering reliability but also social equity, cyber-physical risks, and regulatory compliance. This gap highlights the need for a comprehensive review that bridges these technical, social, and regulatory dimensions within a unified community-centric framework.

Motivated by these challenges, this paper presents a comprehensive review of community-centric resilience enhancement strategies, with a focus on critical infrastructure interdependencies, AI- and ML-enabled approaches, and associated regulatory frameworks. This integrated
perspective provides a holistic synthesis for understanding how technical resilience
strategies interact with societal impacts and regulatory environments in modern
energy systems.

The remainder of the paper is organized as follows. 
Section~\ref{sec:enhancementStrat} reviews resilience enhancement strategies for power systems under extreme events. 
Section~\ref{sec:AI} presents recent advances in applying AI and ML techniques to resilience analysis and enhancement. 
Section~\ref{sec:com_and_PS} discusses the interdependence between power system resilience and community resilience. 
Section~\ref{sec:Law} reviews techno-legal and regulatory considerations relevant to resilient power systems and communities. 
Finally, Section~\ref{sec:conclusion}
summarizes the main findings, identifies research gaps, discusses limitations and opportunities for advancing community-centric power system resilience, and proposes directions for future work. 

\section{Power Systems Resilience Enhancement Strategies}\label{sec:enhancementStrat}
Resilience assessment provides quantitative methods to identify vulnerabilities of power systems, enabling proactive strategies to strengthen resilience before extreme events occur.
Here, different types of resilience-enhancing strategies are identified and explained.
%
\subsection{Hardening Power Network}
System hardening refers to modifications made to power system infrastructure to minimize vulnerability to severe events \cite{mishra2021review,bie2017battling}. 
Hardening is particularly effective at mitigating the impact of extreme events in distribution systems. 
Typical hardening techniques include reinforcing substations, upgrading or undergrounding power lines, and armoring poles \cite{wang2022systematic}. 
The U.S. National Electric Safety Code requires the replacement of poles when their structural ability falls below two-thirds of their original strength 
\cite{wang2022systematic}. 

In \cite{lin2018tri}, a tri-level optimization is proposed that focuses on hardening power networks to enhance the resilience of distribution systems.
It explores strategic decision-making for system hardening through scenario planning under maximum-damage conditions.
System hardening, while serving as an effective resilience enhancement strategy, involves a significant financial burden. 
Consequently, substantial research has focused on developing cost-effective hardening solutions without compromising safety or efficiency \cite{mishra2021review}. In addition to general infrastructure reinforcement, several studies have examined hardening strategies specifically aimed at mitigating lightning-induced failures in transmission systems operating in regions with high lightning density. Lightning strikes are a major cause of transmission line outages. To reduce these risks, mitigation measures such as improving tower-footing grounding resistance, installing additional shield wires or underbuilt ground wires, and deploying transmission-line surge arresters (TLSAs) have been widely investigated. Recent studies have proposed optimization-based frameworks to determine the optimal placement of TLSAs along transmission corridors. For example, several works validated these strategies using realistic transmission-line models in lightning-prone regions such as Brazil and Malaysia, demonstrating significant reductions in lightning-induced outage rates when protection devices are strategically deployed  \cite{Visacro2022LightningReview,Visacro2021CombinedMeasures,Visacro2020TLSAConstraints,Castro2022OptimalTLSAPlacement,Ahmed2023ArresterProtocol}. 
In parallel, risk-based resilience frameworks have been developed to quantify the probabilistic impact of thunderstorms on transmission infrastructure by combining lightning detection data, line outage records, and fragility models. These approaches enable utilities to prioritize protection upgrades and maintenance actions based on lightning exposure and asset vulnerability, with 
applications demonstrated using transmission outage
data from Southwest China and probabilistic analyses calibrated to England's transmission networks 
\cite{Bao2021ThunderstormFragility,Souto2023ProbabilisticLightningImpact}. Beyond these infrastructure-level mitigation strategies,
the integration of distributed energy resources (DERs) into modern power systems introduces additional challenges. 
The uncertainty associated with renewable energy operation complicates traditional scenario-based hardening techniques, as it increases the 
number and severity of system contingencies and potential worst-case scenarios 
\cite{wang2022systematic}.

\subsection{Resource Allocation}
Resource allocation is a resilience enhancement approach that involves strategically deploying DERs at key nodes in the power system to bolster resilience \cite{arghandeh2014local}. 
Given the uncertainty of DERs operation during extreme weather conditions, it is essential to back up the lost power to maintain system stability. This helps strengthen grid resilience while maintaining reliable operation during both normal and extreme conditions \cite{venkateswaran2020approaches}. 
Strategically positioning DERs within the network can reduce the need for critical load curtailment and enhance overall system resilience \cite{dharmasena2022algorithmic}. 
%
%
In \cite{yang2022optimal}, a two-stage stochastic mixed-integer programming model was developed to optimize the location and sizing of DERs (in the first stage), minimize the operating costs of DERs and load shedding (in the second stage). 
Other design decisions, like optimal DER size, can be made using similar optimizations with the objective to maintain resilience above a specified threshold \cite{wang2021three}.
While optimization-based approaches can determine the optimal placement and sizing of DERs under predefined conditions, the machine learning methods reviewed in Section 3 provide data-driven support for resource allocation by leveraging the increasing availability of large-scale data from monitoring systems and intelligent devices \cite{xie2020review}.

\subsection{Optimal Scheduling of Repair Crews}
After an extreme incident, quick repairs of the power system are needed to minimize load-shedding losses. 
Maintaining resilience requires a proactive repair crew deployment approach to prepare for post-disaster repair and restoration. 
Optimization models have been proposed to minimize load shedding by sending repair crews to the ideal locations before extreme events \cite{bian2021proactive}. 
Their goal is to maximize the expected restored load while minimizing the repair duration.
In \cite{arif2017power}, Anmar Arif et al. propose a MILP based co-optimization of repairs, reconfiguration, and DER dispatch in a two-stage method for outage management. 
However, these optimization problems operate under assumptions of full system observability, known damage states, and possible actions. 
As reviewed in Section 3, deep reinforcement learning methods in references \cite{kamruzzaman2021deep} and \cite{zhao2022deep} learn adaptive policies from state–action–reward interactions that generalize across system states and scale to large networks without relying on predefined scenario sets.

\subsection{Network Reconfiguration}
Network reconfiguration refers to the modification of the network topology by opening normally closed sectionalizing switches and closing normally open tie switches.
In order to follow a set of operation constraints, network reconfiguration problems are typically modeled using various objectives, like minimum switching operations, operation costs, or power loss \cite{song2016milp}. 
A number of technologies are employed to improve reconfiguration capacity and distribution system resilience, including smart transformers, DERs, fault prevention and detection systems \cite{mishra2021review}. 
Network reconfiguration can be particularly effective at improving power system resilience when more sophisticated measuring instruments are used.
Devices like distributed remote controllers, phasor measurement units (PMUs), communicating relays, reclosers, advanced supervisory control and data acquisition (SCADA), and outage management systems (OMS) allow for remote reconfiguration, which can reduce load interruption times. 
A more effective approach uses the real-time status of the system during each stage of an event to guide subsequent decisions \cite{wang2022systematic}. In Section 3, it will be discussed that AI and machine learning methods enable more adaptive and real-time network reconfiguration by learning control policies that map system states to switching actions. 
  
\subsection{Protection Strategies for Resilient Grid Operation}

Protection systems represent a critical operational layer for maintaining resilient grid performance during faults and extreme disturbances. Traditional protection schemes rely on fixed relay settings designed for relatively stable operating conditions. However, increasing penetration of DERs, bidirectional power flows, and dynamic network configurations have reduced the effectiveness of conventional protection coordination \cite{11225525}. Consequently, recent research has focused on adaptive and communication-assisted protection strategies capable of responding to real-time system conditions and limiting disturbance propagation.

One emerging direction involves adaptive relay coordination, where protection settings are dynamically adjusted to accommodate changing operating conditions. For instance, a dual-setting protection strategy for microgrids integrates directional overcurrent relays with communication-assisted coordination to improve relay selectivity and response time \cite{Alasali2023MicrogridOCR}. The proposed scheme demonstrated substantial improvements in fault isolation performance. Similarly, adaptive relay frameworks have been proposed to modify the time-multiplier settings of overcurrent relays in real time based on fault impedance variations and operating conditions \cite{Wadie2023AdaptiveOCR}.

Beyond local relay adaptation, wide-area protection systems have been investigated as a complementary mechanism for improving system-level resilience. These approaches utilize synchronized measurements from PMUs to detect faults and evaluate protection system performance across multiple network locations \cite{Yu2019WideAreaPMU}. Simulation studies showed that the method can detect faults within approximately 100 ms and rapidly determine whether primary protection has operated correctly.

Another important class of resilience-oriented protection strategies involves adaptive remedial action schemes (RAS). Unlike conventional RAS implementations with fixed activation conditions, recent studies have explored optimization-based frameworks capable of updating RAS triggers and corrective actions in real time. Optimization-driven RAS design has been shown to significantly reduce post-contingency constraint violations by considering multiple operating scenarios and dynamically adjusting corrective actions under changing system conditions \cite{Rangarajan2024AdaptiveRAS}.

Communication-enabled protection strategies have also gained attention as a cost-effective method for improving protection selectivity in DER-dominated distribution networks. A recent study introduced a hybrid overcurrent protection scheme that integrates conventional relays with communication-based angle comparison logic \cite{Banerjee2024HybridProtection}.

\begin{figure}[t!]
    \centering
    \includegraphics[width=.85\textwidth]{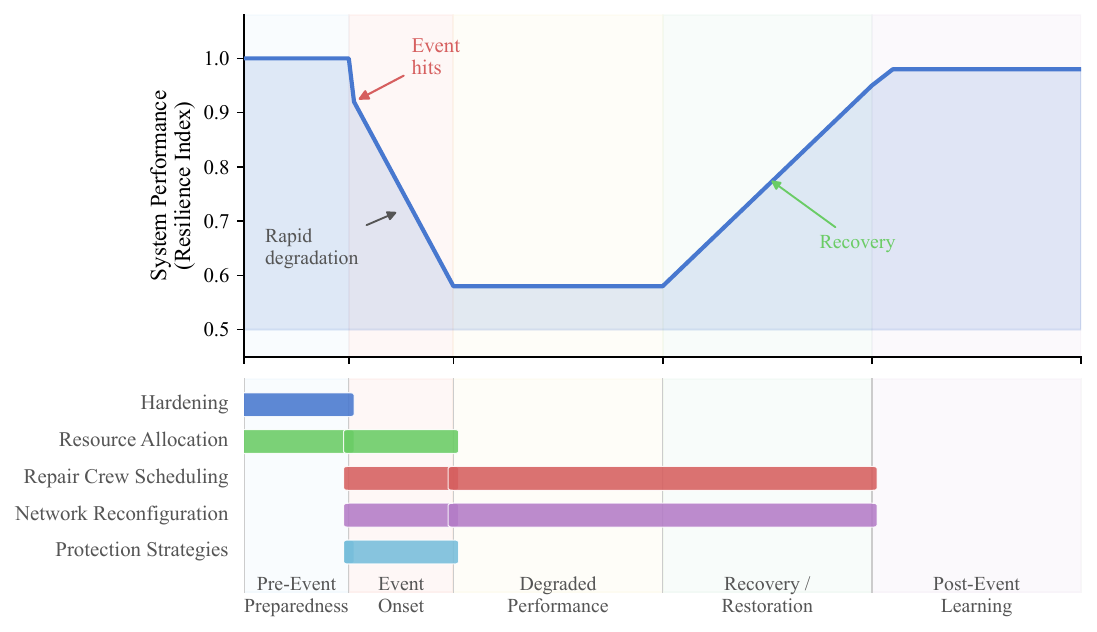}
    \caption{Power system resilience enhancement strategies to the extreme event lifecycle.}
    \label{enchancement lifecycle}
    \vspace{-1em}
\end{figure}

The resilience enhancement strategies presented in this section can be understood within a unified temporal framework of system response to extreme events. Fig. \ref{enchancement lifecycle} shows the extreme event lifecycle and positions each strategy according to its operational role across pre-event preparedness, event onset, degradation, and recovery phases. Hardening measures are exclusively pre-event investments, while resource allocation spans the pre-event and onset phases; protection strategies activate at fault onset to limit degradation; and network reconfiguration and repair crew scheduling operate continuously from event onset through full restoration.

\section{AI Methods for Power Systems Resilience}\label{sec:AI}

\begin{figure}[t!]
    \centering
    \includegraphics[width=.85\textwidth]{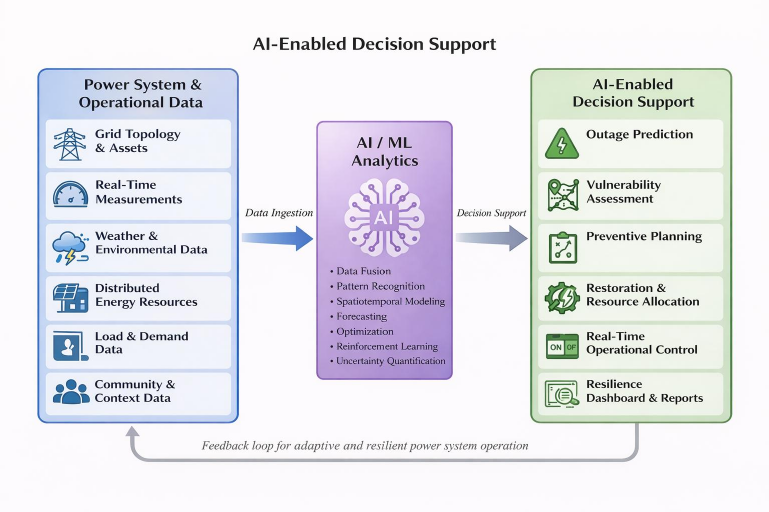}
    \caption{AI-enabled decision support framework for power system resilience.}
    \label{ai_power}
    \vspace{-1em}
\end{figure}

AI and ML methods are emerging as powerful power system operation and planning tools applied to a variety of tasks, particularly to improve the effectiveness of the resilience enhancement methods discussed in Section 2 \cite{fatehi2023machine,fatehi2023ai,xie2020review,11272230,KimAdeleke2025AMI}. Fig. \ref{ai_power} illustrates the integration of heterogeneous datasets into analytics modules for spatiotemporal modeling, prediction, and optimization, enabling decision support such as resilience enhancement. 

Most ML techniques can be categorized into four groups: 1) Traditional ML, 2) Artificial Neural
Network (ANN), 3) Reinforcement Learning (RL), and 4) Advanced Generative and Language Models. Table \ref{Table:AI} summarizes different ML approaches for power system resilience, highlighting their performance, limitations, and related extreme events. The variation in reported accuracy across studies is primarily attributed to differences in evaluation metrics, nature of considering extreme events, and prediction horizon. High accuracy (e.g., $>$90\%) often results from simplified classification tasks (0-no outage, 1-outage), or event-specific (e.g., hurricane, lightning) outage prediction while lower accuracy ($<$80\%) typically reflect longer prediction horizons, and high resolution (e.g., hourly, daily). 

Power outage prediction has emerged as a critical area of research to bolster the resilience of energy systems against various disruptions, including weather events, equipment malfunctions, and grid instabilities.
Logistic regression is a simple, fast, and robust method for outage prediction \cite{eskandarpour2016machine}. 
Higher-order and weighted logistic regression improve performance by capturing nonlinear relationships. Some techniques offer higher accuracy than logistic regression. For instance, Support Vector Machines (SVM) has been utilized to predict outages by assessing the condition of grid components \cite{eskandarpour2017leveraging}. 
%
One other notable method is to use ensemble learning techniques. 
For example, ADABOOST+ has been used to analyze the impacts of lightning and wind on power outages across four cities in Kansas \cite{kankanala2013adaboost}.  
In \cite{yang2020enhancing}, the authors enhance prediction model accuracy by pre-classifying events into three severity levels using Quantile Weight Distance (QWD): low ($<$ 100 outages), moderate (100–1,000 outages), and high ($>$ 1,000 outages). 
A conditioned outage prediction model (OPM) is then developed using a Bayesian Additive Regression Tree (BART), Random Forest (RF), and Ensemble Decision Tree (ENS) to outperforms the standard OPM. 
 
While traditional machine learning models are limited in capturing complex nonlinear relationships, ANNs address this limitation, with additional hidden layers. In \cite{udeh2022autoregressive}, Convolutional Neural Networks (CNN) combined with long-short-term memory (LSTM) have been applied to predict outages across selected counties in New York, studying the socioeconomic impacts on vulnerable populations based on household income and racial demographics. In many cases, model performance is improved through effective data preprocessing, such as synthetic minority over-sampling technique (SMOTE) \cite{watson2022improved}.
In \cite{wang2024deep},
the role of socioeconomic variables in power outage prediction is also examined.

\begin{table}[t!]
\centering
\caption{Summary of machine learning in different studies to enhance power resilience}
\label{Table:AI}

\resizebox{1\textwidth}{!}{   
\renewcommand{\arraystretch}{1.25}
\small

\begin{tabular}{p{2.5cm} p{3cm} p{8cm} p{3.5cm} p{4cm}}
\toprule
\textbf{Category} &
\textbf{Methodology} &
\textbf{Key Findings} &
\textbf{Drawbacks} &
\textbf{Extreme Event} \\
\midrule

\textbf{Traditional ML} &
Regression-based Model \cite{kankanala2011regression,eskandarpour2016machine,sharma2023forecasting,arora2024quasi} &
Simple or linear regression accuracy typically 60--70\%, where higher-order and weighted logistic regression can exceed 90\%.
  Event-based (customer affected): 87-93\% \cite{arora2024quasi,eskandarpour2016machine}, hourly (binary classification of outage probability): 60-75\% \cite{sharma2023forecasting}, daily (outage counts): 60-75\% \cite{kankanala2011regression} &
Tend to underestimate high-impact low probability events &
Wind and lightning events \cite{kankanala2011regression}, Hurricanes (simulated) \cite{eskandarpour2016machine}, Hurricane Ian (Florida, 2022)\cite{arora2024quasi}, Extreme weather outages \cite{sharma2023forecasting} \\

 &
Ensemble Learning \cite{kankanala2013adaboost,madasthu2023ensemble,goforth2022rapidity} & Predictive accuracy ranging between 65--99\%; boosted ensemble achieves higher accuracy by adaptive weighting. Daily (outage counts): 65-80\% \cite{kankanala2013adaboost,madasthu2023ensemble}, event-based (binary classification): 95-99\% \cite{goforth2022rapidity}. 
  & Model lacks explicit representation of network topology & Wind and lightning events \cite{kankanala2013adaboost}, Hurricanes and storms (North Carolina)\cite{madasthu2023ensemble} \\
\midrule

\textbf{Artificial Neural Network (ANN)} &
Classical ANN \cite{kankanala2012estimation,onaolapo2022comparative} & High predictive accuracy ($>$80\%), outperforming regression due to nonlinear mapping ability. Daily (outage count): $\approx 80\%$, seasonal (outage count): 99\%. \cite{onaolapo2022comparative}. & Prone to overfitting with limited data & Wind and lightning (Kansas)\cite{kankanala2012estimation}, Rainfall, storms \cite{onaolapo2022comparative} \\

 &
Deep learning (CNN, LSTM, GNN) \cite{udeh2022autoregressive,gautam2023transductive,wang2024deep} & Achieved 90--99\% accuracy; adding additional hidden layer improves model accuracy. Hourly (customer affected): 91-95\% \cite{udeh2022autoregressive,wang2024deep}, Event-based (binary classification): 92-99\% \cite{gautam2020resilience}.  & Requires large, diverse
datasets & Hurricanes, floods (New York)\cite{udeh2022autoregressive}, High winds and Ice-storms (Michigan)\cite{wang2024deep} \\
\midrule

\textbf{Reinforcement Learning (RL)} &
Classical RL \cite{wang2023towards,li2021integrating,abdelmalak2022network} & Effective for system restoration, control, and decision making, serving as an action-based framework that emphasizes optimal decision, however, its application to outage prediction remains limited. & Use small-case network & Hurricanes,  Storms \cite{li2021integrating,wang2023towards,abdelmalak2022network} \\

 &
Deep RL \cite{kamruzzaman2021deep,zhao2022deep,gautam2023postdisaster,corrado2023deep} & Integrates neural networks to learn optimal actions in large networks where classical RL struggles with high-dimensional states. Resilience indices increase by 5--15\% and reduce computation time. & Computationally expensive and less interpretable & Natural disaster\cite{zhao2022deep,gautam2023postdisaster}, Cyber-attack \cite{corrado2023deep}, Hurricane \cite{kamruzzaman2021deep} \\
\midrule

\textbf{Advanced
Generative
and Language
Models} &
Transformer-based \cite{he2024multi,yao2025ai} & A type of neural network architecture that process data in parallel rather than sequentially. Reduce forecast errors compared with numerical models. & Dependence on multi-source data availability and integration & Natural disaster, cyber-attack. \\

 &
LLM-based Models \cite{zhao2026large,alqudah2023enhancing,yao2025ai} & An application built on Transformer architecture, accuracy can be improved through parameter tuning and contextual adaptability.  &  High computational and storage cost & Hurricanes, cyber-attacks, natural disasters. \\
\bottomrule
\end{tabular}
}
\end{table}

Reinforcement learning (RL) has emerged as a prominent technique in stability control and restoration tasks.  
In RL applications, the environment is modeled as a finite-horizon Markov decision process, defined by the tuple $(S,A,p,r,d_0,\gamma)$. 
Here, $S$ and $A$ represent the state and action spaces, respectively.
$p(s' \mid s,a)$ is the state transition function indicating the probability of transitioning to state $s'$ from state $s$ after taking action $a$; $r(s,a)$ is the reward function for taking action $a$ in state $s$; $\gamma$ is the discount factor for future rewards, and $d_0$ is the initial state distribution 
\cite{corrado2023deep}.

Different RL tasks are characterized by unique state and action spaces, along with task-specific reward functions focused on using RL for microgrid (MG) formation post-disaster 
\cite{zhao2022deep,gautam2022reconfiguration,abdelmalak2022network,gautam2022post,tightiz2021resilience}. 
These models often have discrete action spaces (e.g., toggling switches on and off for specific DERs), and sometimes continuous spaces (e.g., adjusting setpoints for selected DERs) \cite{abdelmalak2022network,corrado2023deep}. In addition, RL has been applied to optimize restoration strategies, such as dispatching mobile energy resources, and managing repair crews \cite{gautam2023postdisaster,wang2023towards}. 
%
Another study discuss scheduling repairs for components and compares the performance of the Q-learning algorithm with Greedy Search (GS) and Exhaustive Search (ES) \cite{li2021integrating}. 


Other ML approaches are also instrumental in enhancing power system resilience. 
Predictive control policies for MG management have employed tree-based models to forecast load demands and solar production \cite{gutierrez2021weather}. 
Bayesian Networks and hybrid statistical algorithms like the Fragility-curve Monte Carlo Simulation Scenario Reduction (FC-MC-SCENRED) have been compared for pre-hurricane system optimization for distribution networks \cite{omogoye2023comparative}.
Graph neural networks (GNNs) have been proposed to assess node criticality and identify vulnerable lines in cascading outage studies \cite{gautam2023transductive,paradell2021increasing}. 
%
For disturbance management, CNNs paired with autoencoders have been developed to dynamically adjust shunt capacitors and transformers during electromagnetic pulse disturbances \cite{zhang2022machine}. 
A one-dimensional CNN has been designed to detect data intrusions in smart meters \cite{yaldiz2023resilience}. Advanced deep learning techniques have been applied for system-wide online load restoration with high wind power penetration \cite{zhao2022deep}, and MG planning in active distribution systems using Bayesian Regularization Backpropagation \cite{vilaisarn2022deep}. 

Recent advancements in transformer-based generative intelligence offer potential to apply in the power systems to better capture complex patterns between parameters. The transformer-based model can tune millions of parameters at a time \cite{yao2025ai}. Such large-scale data-driven modeling capabilities can also support emerging digital twin frameworks for community-scale energy systems, where integrated models simulate infrastructure behavior, outage impacts, and resilience interventions across interconnected urban systems \cite{Shittu2023DigitalTwinSubstations}. The models, such as eGridGPT, are used to train and forecast more accurate power grid analysis and decision making \cite{choi2024generative}. Domain-specific large language
model (LLM) framework extends beyond conventional predictive learning by introducing reasoning, context awareness, and multi-modal understanding of the grid’s characteristics \cite{yao2025ai}. Self-attention mechanisms can effectively combine information from different sources and understand how events change over time and locations \cite{he2024multi,alqudah2023enhancing}.  Another model predicts outage occurrence as a binary classification problem and is evaluated using Precision, Recall and Area Under the
Precision-Recall Curve (AU-PRC), enabling to predict up to several hour ahead of the event \cite{alqudah2023enhancing}. However, the model requires high-performance computing infrastructure to process datasets for training and real-time inference. 

Despite advantages of AI methods, several challenges remain for the real-world deployment, including issues related to data privacy, adaptability to dynamic conditions, and broader ethical considerations\cite{zhao2026large}. For instance, AI models may generate inaccurate or fabricated outputs (hallucinations) due to issues in training data quality, bias, and inconsistencies between training and deployment conditions \cite{yao2025ai}. Additionally, due to the data accessibility and AI's black-box nature, they are vulnerable to attacks such as backdoor manipulation and data exploitation. 

In addition, sensitive information such as customer consumption patterns and infrastructure data cannot be freely exchanged due to compliance requirements under regulatory frameworks such as the EU General Data Protection Regulation (GDPR) \cite{eu_gdpr_regulation_2016_679} and the North American Electric Reliability Corporation Critical Infrastructure Protection (NERC CIP)\cite{NERC_CIP_Standards}. These constraints lead to fragmented data silos and reduced model effectiveness. To address these limitations, federated learning (FL) has emerged as a distributed machine learning framework that enables collaborative model training without sharing raw data. Instead, local models are trained at individual data sources and only model parameters are exchanged  \cite{zheng2024advancing}. This approach improves scalability and supports learning across geographically distributed and heterogeneous power system environments. However, challenges such as communication overhead, non-identical data distributions, security vulnerabilities, and aggregation complexity can affect model performance \cite{cheng2022review}.

\vspace{-0.75em}

\section{Resilience of Interdependent Power and Community Infrastructure Systems}\label{sec:com_and_PS}

Modern power systems operate within an interconnected ecosystem of critical infrastructures whose joint performance determines overall resilience.
While power system resilience traditionally refers to the grid’s ability to withstand and rapidly recover from HILP events \cite{shao2017integrated}, community functionality and broader infrastructure recovery are inherently linked to this capability.
Community resilience—defined as the capacity of populations and institutions to prepare for, absorb, and recover from disasters—depends heavily on the continuity of electricity supply and the robustness of interdependent systems such as transportation, water, gas, and telecommunication networks \cite{zhang2024seismic}.
In highly electrified and infrastructure-reliant societies, disruptions in the power grid can cascade across sectors, amplifying socioeconomic impacts.
Accordingly, assessing resilience requires a system-of-systems (SoS) perspective that captures both technical inter-dependencies and community consequences. This perspective highlights that resilience outcomes are not determined solely by the reliability of the power grid itself, but also by how disruptions propagate through interconnected infrastructures and affect communities that depend on these services.

\subsection{Infrastructure Interdependence and System-of-Systems Perspective}
\begin{figure}[t!]
\vspace{-0.5em}
  \centering
  \includegraphics[width=0.8\textwidth]{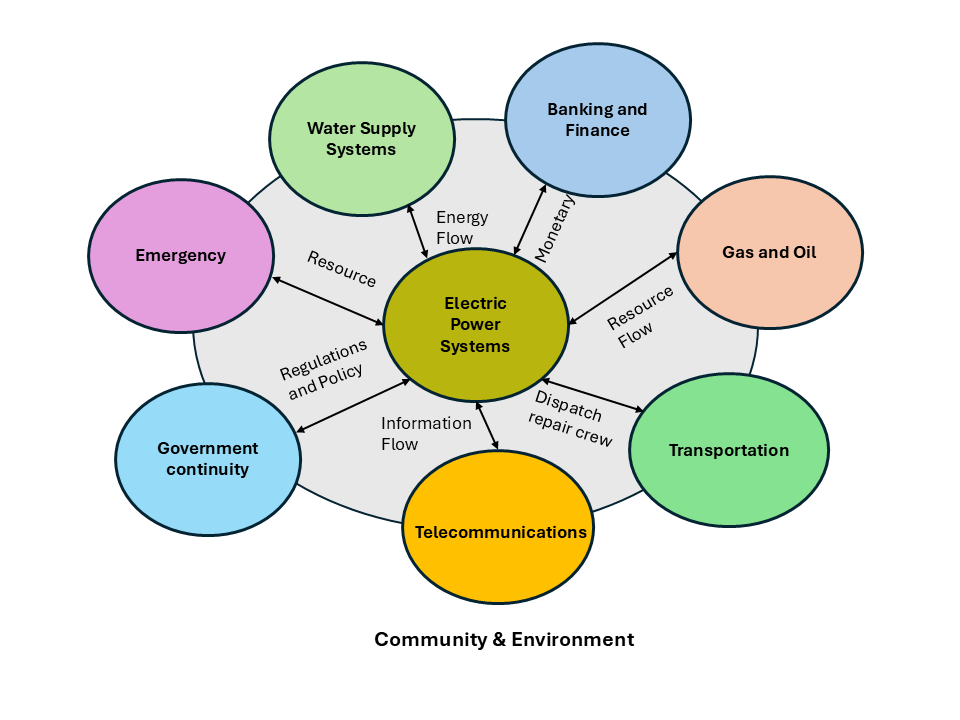}
  \captionsetup{font=small, justification=centering}
  \caption{Interdependency of critical infrastructure systems as a SoS.}
  \label{fig:CI1}
\end{figure}

The electric grid is tightly coupled with other critical infrastructures (Fig.~\ref{fig:CI1}). Under normal conditions, such coupling enhances system efficiency through coordinated control, information sharing, and optimized resource allocation. However, during HILP events, these same inter-dependencies accelerate disruption propagation and amplify cascading failures across sectors 
\cite{paul2021vulnerability,mohamed2019rising,CI2025,brunner2024understanding}. 

From a resilience perspective, these coupled infrastructures represent both stabilizing supports and potential pathways for systemic collapse. Telecommunications enable situational awareness and remote control; gas and water networks supply essential fuels and cooling resources for generation; and transportation systems facilitate repair crew mobilization, logistics, and power system recovery activities \cite{mahzarnia2020review}. 
Recent studies highlight this need by quantifying cross-sectoral sensitivity, recovery delays, and performance degradation under multiple hazards. Table~\ref{Table:A} summarizes representative studies examining the interdependence among the electric power systems and other communities and critical infrastructures.

Water infrastructure exhibits strong dependence on electricity for pumping and pressure maintenance. Optimization-based resilience modeling in \cite{9707888} showed that placing DERs near pump stations can increase Water Distribution System (WDS) resilience by 13.35\%. Maintaining pump operation during outages is particularly important because even short power interruptions can disrupt water pressure, sanitation, and emergency response capabilities within affected communities.
Under more severe disruptions, potable water systems experience drastic degradation when electrical service is lost. Using the Water Network Tool for Resilience (WNTR) with Pressure-Dependent Demand modeling, \cite{portable} demonstrated that hurricane-induced power outages reduce potable water system resilience by nearly 50\%. 
Such performance losses illustrate how electrical failures can quickly propagate into essential public health services.
 
Dynamic optimal energy flow modeling in \cite{sun2023resilience} showed that incorporating gas storage improves gas system resilience by 29.5\%. 
and that adding backup generators increases heating-system resilience by 15.6\% following hurricanes, indicating that cross-sector energy resources can provide critical redundancy during extreme events. 
Multi-hazard Monte Carlo simulations in \cite{RAVADANEGH2022103687} further show that wind and earthquake events reduce energy resilience by approximately 7–14\%. These cascading effects can significantly extend recovery timelines for communities reliant on multiple energy services.

Transportation infrastructure also relies heavily on the power grid for operational continuity. Seismic resilience simulations in \cite{zhang2024seismic} showed that increasing earthquake magnitude from $M=6$ to $M=8$ reduces transportation network resilience by nearly 50\%. In large-scale disasters, reduced mobility can therefore prolong outages and slow community recovery processes. 
Complementary robust optimization studies in \cite{wang2018resilience} demonstrate that strategic hardening of critical power lines and placement of DERs significantly reduce disruption costs and preserve the operability of transportation-critical loads such as traffic signals during HILP events.

Linear regression analysis of post-hurricane recovery in \cite{infrastructures9110208} quantified strong cross-sector dependencies during restoration. Water, telecommunications, and hospital systems relied on power recovery for 89\%, 93\%, and 77\% of their restoration trajectories, respectively. Delays in power restoration, therefore, tend to propagate across multiple infrastructures, extending the overall duration of community disruption following extreme events.

These studies indicate that disruptions in the electric grid can propagate across multiple sectors, amplifying societal impacts and delaying community recovery. This SoS perspective suggests that resilience planning should prioritize not only electrical infrastructure reliability but also the protection of services that enable essential community functions such as water supply, transportation, healthcare, and communications.


\subsection{Resilience Hubs for Community Infrastructure Support}\label{sec:ResilienceHubs}
Grid resilience strategies are most meaningful when they translate into tangible benefits for the communities they serve. Resilience hubs represent this translation, as they
 provide a comprehensive, community-centered approach to addressing interdependent, multifaceted vulnerabilities. 
They recognize the interconnected nature of critical community infrastructure systems and the socioeconomic dimensions through which HILP events affect community members. 
Resilience hubs integrate public health and safety goals with technical considerations for strengthening power, transportation, communications, and essential services. 

\begin{figure}[ht]
\vspace{-0.2em}
  \centering
  \includegraphics[width=0.78\textwidth]{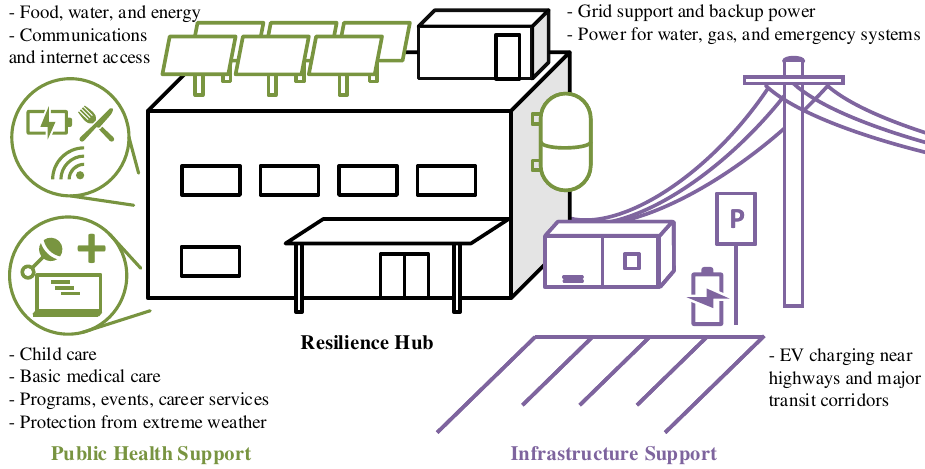}
  \captionsetup{font=small, justification=centering}
  \caption{Schematics of a Resilience Hub.}
  \label{fig:RHub}
\end{figure}

Resilience hubs are community-oriented facilities, such as schools, community centers, or libraries, that provide everyday public services during normal operating conditions and serve as disaster shelters, resource centers, and infrastructure support during emergencies \cite{baja2018} (Fig. \ref{fig:RHub}). 
\

Resilience hubs reflect the interdependence of community and grid resilience, addressing both immediate individual social vulnerability concerns and technical system-level infrastructure shortfalls \cite{farley2024}. When strategically designed as integrated techno-physical-social system nodes, they can serve as a foundation for community and grid resilience.  
For example, a resilience hub may focus on supporting medically vulnerable populations, stabilizing public health services, coordinating emergency response or recovery efforts, ensuring access to cooling or clean water, or enhancing grid reliability using DER and adaptive load management.  

Resilience hubs are equipped with resilient backup power systems that not only ensure uninterrupted on-site access to critical services but can also provide grid-supporting functions \cite{farley2024}.  
This flexible capacity can stabilize nearby interdependent community infrastructure by supporting water pumping stations, emergency communications, transportation assets, and reliable electric vehicle (EV) charging during outages or evacuations.  
As a result, resilience hubs can be strategically located either to optimize accessibility for residents \cite{farley2025} or near critical infrastructure during grid disruptions \cite{rodriguezroman2024}.  

By forming a resilience hub-centered MG using smart switching and grid reconfiguration techniques described in Section \ref{sec:enhancementStrat}, operators can support critical infrastructure and services (such as cooling during extreme heat events) during cascading grid failures.
Resilience hubs can also coordinate with grid operators to support broader black start recovery. 
The RL-based reconfiguration methods and outage prediction models discussed in Section \ref{sec:AI} further enhance this capability.

The framework in \cite{farley2025} designs and evaluates energy systems for resilience hubs in a region prone to extreme heat, identifying communities with limited access to air conditioning and heightened vulnerabilities to power outages.  
The resulting resilience hub design incorporates a DER portfolio sized to maintain safe indoor temperatures during a heatwave for up to four days without relying on the grid.  
Similarly, the grid-aware optimization in \cite{gautam2024} demonstrates how hub-centered MGs can mitigate outage impacts by balancing local resilience benefits with broader system constraints. 
Outage prediction models can guide optimal hub siting by identifying where outage risk, social vulnerability, and infrastructure interdependence converge, while RL-based frameworks can dynamically dispatch emergency responders, manage EV charging, and allocate backup generation in response to evolving event conditions.
These examples illustrate how resilience hubs are not merely a community amenity, but the point at which grid resilience technologies, AI-driven prediction, and community vulnerability considerations converge into tangible support for critical infrastructure.

\begin{table*}[t!]
\centering
\caption{Summary of the interconnection between  community resilience and power systems resilience.}
\label{Table:A}

\resizebox{1\textwidth}{!}{   
\renewcommand{\arraystretch}{1.25}
\small

\begin{tabular}{p{3.5cm} p{4cm} p{4cm} p{8.0cm} p{3cm}}
\toprule
\textbf{Category} & 
\textbf{Community / Infrastructure} & 
\textbf{Methodology} & 
\textbf{Key Findings} & 
\textbf{Extreme Event} \\
\midrule

\textbf{Behavioral\cite{loni2024data,dominianni2018power,pgad295,Rao2025,Do2023,ANDRESEN2024103573}} 
& Older adults, Low-income households, DME users  
& Telephone survey \cite{dominianni2018power} 
& 58\% reported outage preparedness; 70\% of DME users were prepared, but 74\% had serious health concerns during outages 
& General climate change \\

\midrule

& Home EV chargers, Educational attainments, Low income residents 
& SoVI analysis using PCA, ARF, VBSA, Sobol \cite{loni2024data} 
& Most influential factors affecting outage vulnerability: Education (\%99.7) — higher levels decrease vulnerability; Affected Customers (\%86.8) — larger outage exposure increases vulnerability
& Heatwaves, storms \\

\midrule

& Low-income, elderly, socially vulnerable  
& SDM regression using SoVI \cite{pgad295} 
& 0.1 increase in socioeconomic vulnerability results in a 6.1\% increase in outage duration; Counties with the highest delays in power restoration—+6 hours longer on average—had very high SoVI = 0.91  
& Hurricane \\

\midrule

& High racial/ethnic vulnerability 
& Multilevel Bayesian modeling using gamma regression \cite{Rao2025} 
& Counties with SoVI $>0.9$ faced the longest outage duration
& Hurricanes, heat waves, floods, winter storms \\

\midrule

& Urban centers, rural counties 
& SoVI analysis using spatial cluster, and non-parametric statistical tests\cite{Do2023} 
& High-SoVI counties experienced approximately $2\times$ more outages lasting $\geq 1$ hour and $3\times$ more outages lasting $\geq 8$ hours compared to low-SoVI counties 
& General climate change \\

\midrule

& Diverse urban households 
& Household surveys using Kruskal–Wallis tests\cite{ANDRESEN2024103573} 
& Minority households reported longer outages: 58.1\% of minority respondents compared to 40.4\% of white respondents; Low-income households experienced more frequent outages, with 17.2\% versus 8.4\% among higher-income households 
& General climate change \\
\midrule

\textbf{Infrastructure\cite{9707888,sun2023resilience,portable,zhang2024seismic,RAVADANEGH2022103687,infrastructures9110208}}
& Water distribution systems 
& Resilience metrics computation using SOCP optimization \cite{9707888} 
& Placing DGs near pumps increased RA by approximately 13.35\%, indicating stronger WDS performance relative to PDS degradation; Enabling local backup generation increased RA by about 3.58\%, indicating a modest improvement in maintaining WDS performance 
& Storm \\

\midrule

& Natural Gas, and Heating System 
&Dynamic optimal energy flow modeling with resilience metrics\cite{sun2023resilience} 
& Gas system resilience improved by 29.5\%— when gas storage capacity was considered; Heat system resilience increased by 15.6\%—after adding backup generators 
& Hurricane \\

\midrule

& Potable water system 
&  WNTR tool with Pressure-Dependent Demand \cite{portable}  
& Power outages reduced potable water system resilience by 50\% (e.g., water service dropped below 50\% within 8.8 days), and extended full recovery time by over 300\% (e.g., from 6.7 to 28+ days) 
& Hurricane \\

\midrule

& Urban transportation (road network) 
& Seismic resilience using Monte Carlo \cite{zhang2024seismic} 
& Resilience drops sharply as earthquake magnitude increases --- at $M = 8$, network resilience is approximately 50\% of its value at $M = 6$, and recovery times are about 15 times longer
& Earthquake \\

\midrule

& Natural gas, and Heating system  
& Monte Carlo simulation of component failures\cite{RAVADANEGH2022103687} 
& Total energy resilience fell from 89\% (wind), 83\% (earthquake), to 75\% (combined) — a 16\% drop under multi-hazards; Energy curtailment reached 76\% under multi-hazard conditions and required 52\% longer restoration time
& Wind Storm, Earthquake \\

\midrule

& Water system, Telecommunications (cell sites), and Hospitals 
& Linear regression analysis of recovery curves\cite{infrastructures9110208}
& Water, telecommunication, and hospital systems rely on power recovery for 89\%, 93\%, and 77\% of their recovery, meaning only 11\%, 7\%, and 23\% of their resilience is independent of the power grid
& Hurricane \\

\bottomrule
\end{tabular}

} 
\end{table*}

\subsection{Socioeconomic and Behavioral Dimensions}
The community-level impact of outages varies widely based on demographic and socioeconomic conditions. Social vulnerability reflects the intersection of economic status, education, age structure, health, and access to critical resources, which collectively determine the capacity of communities to recover from disasters \cite{cutter2003sovi,flanagan2011cdc}. In the context of power systems, social vulnerability determines how technical failures propagate into human and economic consequences: two communities experiencing identical outage durations may face vastly different recovery trajectories depending on their adaptive capacity and institutional support. This makes social vulnerability an essential bridge between grid performance and community resilience \cite{zhang2024seismic,paul2021vulnerability}.

A growing body of research has quantified how socioeconomic and behavioral factors shape outage exposure, duration, and subsequent recovery. A data-driven  Social Vulnerability Index (SoVI) framework in \cite{loni2024data} combined Principal Component Analysis (PCA), Adaptive Random Forests (ARF), and variance-based sensitivity analysis (VBSA) to identify the dominant societal drivers of outage vulnerability. 
Projected SoVI values for 2030 indicate a reduction in vulnerability disparities compared to 2022, reflecting long-term adaptation and incremental infrastructure improvement. Similar disparities were documented in \cite{pgad295}, where a 0.1 increase in socioeconomic vulnerability was associated with a 6.1\% increase in outage duration. Counties with the longest restoration delays (+6 hours above the mean) exhibited very high SoVI values ($\text{SoVI}\approx0.91$). The statistical relationship between SoVI and outage duration highlights how underlying social conditions influence not only the likelihood of disruption but also the pace of infrastructure recovery. Communities with higher vulnerability frequently face constraints such as limited financial resources, aging infrastructure, reduced access to backup energy systems, and weaker institutional support for restoration efforts. Under these circumstances, similar grid disturbances can produce substantially different outage experiences across communities, with vulnerable populations often facing prolonged service interruptions.

Outage exposure also differs across racial and ethnic lines. A multilevel Bayesian analysis in \cite{Rao2025} showed that counties with high racial/ethnic vulnerability \((\text{SoVI} > 0.9\)) experienced the longest outage durations during extreme weather events, including hurricanes, heat waves, and floods. Similarly, a spatially resolved analysis in \cite{Do2023} found that counties with high SoVI experienced approximately two times more outages lasting \(\ge 1\) hour and approximately three times more outages lasting \(\ge 8\) hours compared to low-SoVI counties. These patterns illustrate that outage impacts are not determined solely by the intensity of the hazard or the configuration of the power grid. Instead, pre-existing socioeconomic inequalities influence exposure to disruptions, access to preparedness resources, and the ability of communities to respond effectively when outages occur. As a result, communities with higher vulnerability often experience both greater outage frequency and slower restoration timelines. 

Behavioral factors further amplify these disparities. A survey study in New York City \cite{dominianni2018power} found that although 58\% of respondents reported feeling generally prepared for outages, only 32\% felt capable of enduring multi-day interruptions. Users of durable medical equipment (DME) were more prepared (70\%), yet 74\% still feared serious health complications during outages. 
These findings reveal a crucial gap between perceived and actual preparedness. Additional household-level evidence from \cite{ANDRESEN2024103573} shows that minority households reported significantly longer outages, 
while lower-income households experienced almost twice as many outage events. Such disparities suggest that the societal burden of power outages is unevenly distributed across populations. Differences in preparedness, resource availability, and household resilience capacity can significantly influence how communities experience and recover from disruptions. Table \ref{Table:A} summarizes the interdependency between power systems resilience and SoVI.

Age and health status further influence outage-related risk. Children, older adults, and medically fragile populations experience disproportionately severe impacts from prolonged outages and associated cascading failures \cite{casey2020power,hanke2021renewable}. Documented public health risks include cardiovascular events, respiratory complications, renal failure, heat stress, gastrointestinal illness, and increased mental health burdens as outage durations rise \cite{casey2020power}. Because electricity supports critical services such as medical equipment, climate control, and healthcare infrastructure, prolonged outages can quickly translate into severe health risks for vulnerable populations. To mitigate these outcomes, recent work highlights the importance of community-scale renewable energy resources and medically focused resilience planning to maintain essential electricity access during HILP events \cite{hanke2021renewable}. At the same time, emerging socio-technical models increasingly incorporate behavioral and human-response data  \cite{valinejad2023computational}. Across these studies, a reinforcing feedback mechanism emerges between social vulnerability and power system resilience. Communities with higher vulnerability tend to experience longer outages and slower recovery, which in turn can exacerbate health risks, economic hardship, and reduced access to essential services. These impacts may weaken community preparedness and adaptive capacity, increasing susceptibility to future disruptions and reinforcing existing inequalities.

Recognizing this dynamic has important implications for resilience planning. Infrastructure investments should therefore not rely exclusively on engineering-based risk metrics. Instead, resilience strategies should incorporate equity-centered indicators, when prioritizing resilience investments. Integrating these considerations into grid hardening strategies, DER deployment, and resilience hub planning can help ensure that resilience improvements address the communities most vulnerable to outage impacts.


\section{Techno-Legal Aspects of Resilient Power Systems and Communities}\label{sec:Law}

Community-centric power system resilience is not determined by technical design alone. It also depends on how data are collected, protected, and governed across utilities, healthcare providers, and other critical infrastructures. This becomes particularly important as resilience enhancement strategies increasingly rely on AI-enabled outage prediction, automated restoration support, DER coordination, network reconfiguration, and the operation of resilience hubs. In these settings, legal and regulatory frameworks do not merely constrain implementation after engineering decisions have been made; they directly influence what data may be used, which models are deployable, and how accountability is assigned when failures affect vulnerable communities.

\begin{figure}[t!]
\vspace{-0.5em}
  \centering
  \includegraphics[width=0.85\textwidth]{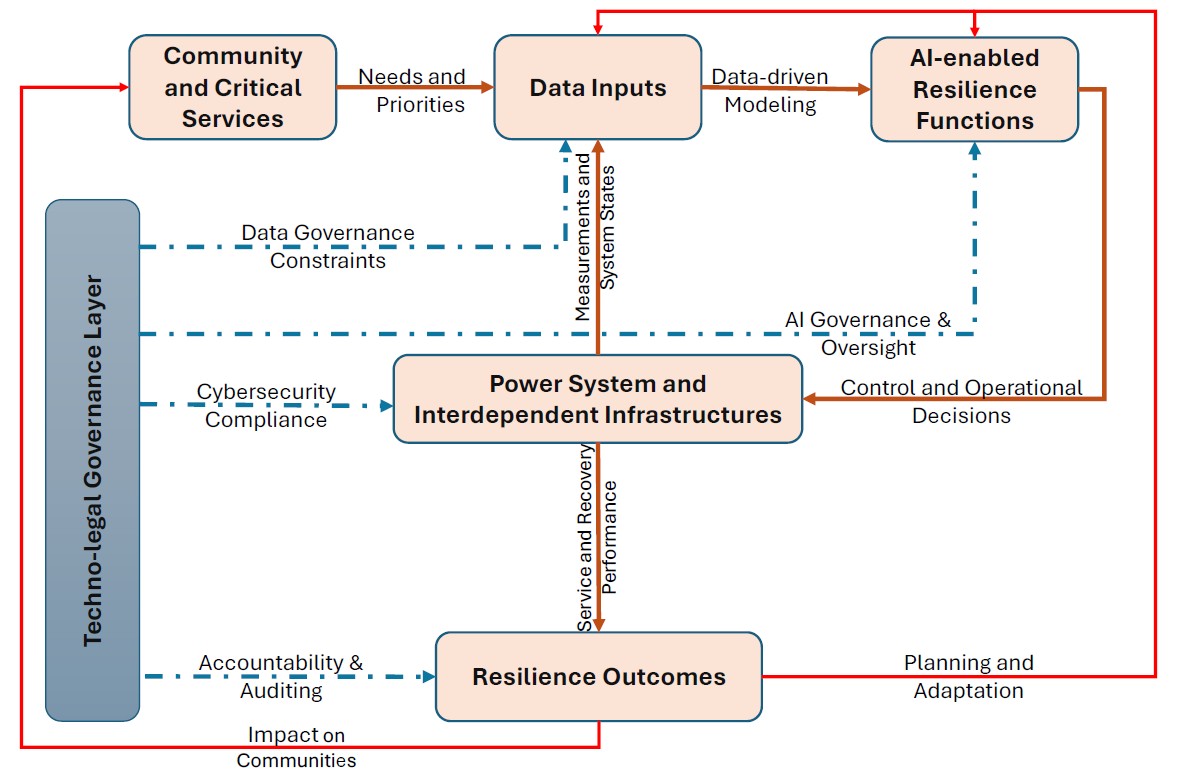}
  \captionsetup{font=small, justification=centering}
  \caption{Techno-legal framework for critical infrastructures as a part of SoS.}
  \label{fig:CI}
\end{figure}

Accordingly, techno-legal governance should be treated as an integral component of community-centric resilience, rather than a downstream compliance layer. As such, resilience evaluation must also account for privacy, cybersecurity, equity, traceability, and lawful coordination across infrastructures and communities.
Figure~\ref{fig:CI} provides a system-level representation of these interactions, illustrating how community needs and critical services
define resilience priorities and drive the collection of diverse data inputs. These inputs include advanced metering infrastructure (AMI) and smart meter data, outage histories, weather and hazard data, topology and asset condition information, and social vulnerability indicators. 
Such data streams enable AI-driven resilience functions.
%

Within this architecture, techno-legal governance mechanisms, such as NERC CIP standards \cite{NERC_CIP_Standards}, the Network and Information Security Directive 2 (NIS2) \cite{eu_nis2_directive_2022_2555}, and data protection regulations including GDPR, the Health Insurance Portability and Accountability Act (HIPAA), and the California Consumer Privacy Act (CCPA)—act as cross-cutting constraints across all layers. These frameworks govern data collection and sharing, regulate AI model design, enforce cybersecurity and operational compliance, and establish accountability, auditing, and liability mechanisms.
These governance requirements arise from the strong interdependencies among critical infrastructures. As a result, resilience must be understood as a coordinated, community-centric objective shaped by infrastructure interdependencies, data-driven decision-making, and regulatory constraints.

\subsection{Data Governance and Privacy Constraints in Resilience Analytics}\label{sec:data_governance}

The AI- and data-driven methods rely on the integration of diverse datasets, including outage histories, grid topology, and weather data. At the community level, resilience-oriented analytics may also require demographic, socioeconomic, and health-related information to identify populations that are likely to experience disproportionate harm during prolonged outages. This creates a central tension for community-centric resilience planning. The models that appear most accurate often depend on fine-grained and heterogeneous data, yet the collection and fusion of such data are restricted by privacy and data governance regimes.

Under the GDPR, the processing of personal data must satisfy principles such as lawfulness, purpose limitation, data minimization \cite{eu_gdpr_regulation_2016_679}. These principles are directly relevant when outage prediction or restoration prioritization models incorporate household-level consumption records, location-linked vulnerability indicators, or data associated with medically dependent users. In the United States, similar concerns arise through sectoral and state-specific frameworks, including HIPAA for health-related information \cite{us_hipaa_1996} and consumer privacy laws such as the California Consumer Privacy Act \cite{california_ccpa_2018}. For community-centric resilience, the practical issue is not only whether data are technically available, but whether their integration is legally justified. 

This tension is especially important for models that aim to identify vulnerable communities or prioritize resilience investments. As discussed in Section~\ref{sec:com_and_PS}, outage impacts vary significantly across communities depending on social vulnerability. 
However, identifying a ``vulnerable'' customer or neighborhood requires the processing of granular data related to health, income, disability, age, or other socially sensitive attributes. This means that resilience planning increasingly falls within the scope of strict privacy and data governance obligations. In practice, the stronger the ambition to make resilience more targeted and equitable, the greater the need for governance mechanisms that lawfully manage sensitive data.

The same issue extends to the AI methods reviewed earlier. High predictive performance alone is not sufficient to justify the operational use of a model in resilience planning. For example, outage prediction systems trained on highly granular household and infrastructure data may improve forecasting performance, but they may also become difficult to reconcile with data minimization requirements, purpose limitation, or restrictions on cross-sector data sharing. In addition, black-box models with large data requirements may reduce transparency regarding data provenance. This concern is particularly salient for advanced transformer-based and LLM-based approaches discussed in Section~\ref{sec:AI}. Although these models may show strong performance in selected prediction tasks, their large computational requirements, opacity, and dependence on extensive training data make them difficult to explain and validate for critical infrastructure use. Their tendency to produce unreliable or hallucinated outputs further limits their suitability for safety-critical and accountability-sensitive settings.

These constraints imply that future resilience analytics should be designed with privacy preservation as a first-order requirement. Techniques such as data aggregation, differential privacy, and federated or distributed learning are promising because they reduce the need to centralize all sensitive data while still enabling collaborative modeling. Similarly, asset and interdependency data, including grid topology, and substation layouts should be treated as sensitive operational information under frameworks such as NERC CIP \cite{NERC_CIP_Standards} and NIS2 \cite{eu_nis2_directive_2022_2555}. In these cases, validated synthetic test systems and controlled data-sharing mechanisms may be preferable for resilience analysis.

\subsection{Cybersecurity by Design for AI-Enabled Resilient Operation}\label{sec:cyber_by_design}

Cybersecurity is not external to resilience engineering. It is a core condition for deploying AI-enabled control and decision-support systems. Several resilience enhancement strategies now depend on digital control, advanced sensing, and learning-based optimization. Examples include repair crew scheduling, MG formation, and network reconfiguration. When these functions are implemented through AI-enabled or data-driven architectures, they inherit not only the benefits of fast adaptation and improved decision support, but also new vulnerabilities related to cyber intrusion, telemetry corruption, and software supply-chain compromise.

This concern is particularly important for RL and deep reinforcement learning (DRL) methods. These methods have been proposed for post-disaster network reconfiguration \cite{zhao2022deep,gautam2022reconfiguration,abdelmalak2022network,gautam2022post}, restoration support \cite{li2021integrating,gautam2023postdisaster,wang2023towards}, and voltage regulation \cite{corrado2023deep}. However, their performance is usually demonstrated under simulated assumptions. In practice, community-centric resilience operations occur under precisely the opposite conditions. Extreme events may degrade communications, distort sensor visibility, interrupt centralized coordination, and create out-of-distribution physical states. Under these conditions, an AI agent may receive manipulated, delayed, or incomplete inputs and may issue unsafe actions.

This creates a direct bridge between resilience enhancement and cybersecurity regulation. NERC CIP \cite{NERC_CIP_Standards} and NIS2 \cite{eu_nis2_directive_2022_2555} both imply that resilience functions must be developed and operated with secure architectures. For AI-enabled resilience functions, they should be assessed for secure sensing, human override, and decision traceability. This is especially important when resilience operations involve PMUs, SCADA, or field devices whose compromise may produce cascading physical effects.

If RL is used for network reconfiguration, then the integrity of topology data and switching status becomes a core safety issue. If AI is used for repair crew scheduling, the model must remain reliable when communication systems are degraded. 
In this sense, the cyber-resilience of AI-driven functions becomes a research problem in its own right. Future work should therefore move beyond proof-of-concept control performance and address formal verification, adversarial robustness, runtime monitoring, secure-by-design requirements \cite{ntia_sbom_minimum_elements_2021}, and human-in-the-loop supervisory control for AI-assisted grid operations.

\subsection{Law-Tech Integration in Interdependent Infrastructures and Resilience Hubs}\label{sec:law_tech_interdep}

As discussed in Section~\ref{sec:com_and_PS}, outages propagate through interdependent infrastructures and create disproportionately severe consequences for socially vulnerable populations. This implies that resilience is not only about asset recovery, but also about ensuring lawful, secure, and coordinated action across institutions and sectors.

This point is especially important for resilience hubs. To function effectively, resilience hubs must be integrated into the wider cyber-physical system of grid operations, restoration prioritization, and interdependent service provision. Restoration and scheduling models can prioritize feeders connected to hubs and medically critical facilities. Network reconfiguration and MG formation strategies can be used to preserve or restore electricity supply to hubs during wider service disruption. 

This interpretation also raises a set of governance questions. Which entities are lawfully permitted to share data on medically dependent populations or socially vulnerable households. Under which legal basis can certain facilities or neighborhoods be prioritized during restoration. How should decision authority, liability, and accountability be allocated when AI-assisted tools are used. These questions show that community-centric resilience depends on institutional preparedness as much as technical preparedness.

For this reason, law-tech integration in interdependent infrastructures should include predefined data-sharing templates, lawful bases for emergency processing under GDPR \cite{eu_gdpr_regulation_2016_679}, and cross-sector emergency exercises that reflect cascading infrastructure risks. Procurement processes for grid and community resilience equipment should embed explicit requirements for NIS2 \cite{eu_nis2_directive_2022_2555}, NERC CIP \cite{NERC_CIP_Standards}, and AI-governance readiness \cite{eu_ai_act_2024}. 
In parallel, techno-legal performance indicators, such as time to detect and report cyber incidents, and completeness of cross-sector data governance documentation, should be tracked alongside traditional resilience metrics \cite{GDPR_DPIA_Art35}.

\subsection{Comparative EU and US Regulations and Their Engineering Implications}\label{sec:comparative_reg}

The EU and US regulatory landscapes both address critical infrastructure resilience, cybersecurity, and AI oversight, but they do so through different legal structures. These differences have direct consequences for the design, procurement, and operational deployment of community-centric resilience solutions.

In the EU, the regulatory environment is becoming increasingly integrated through the combined effects of the GDPR \cite{eu_gdpr_regulation_2016_679}, the NIS2 Directive \cite{eu_nis2_directive_2022_2555}, the CER Directive, the Energy Efficiency Directive \cite{EU_EED_2023}, and the EU AI Act \cite{eu_ai_act_2024}. Together, these instruments create a structured governance environment in which data protection, cybersecurity, and AI oversight are increasingly linked. This has several technical implications. First, resilience-oriented AI systems that affect critical infrastructure operations are likely to face stronger requirements. 
 Second, the stronger emphasis on accountability may favor more interpretable and auditable AI architectures. Third, compliance requirements are likely to shape model documentation, validation workflows, and post-deployment monitoring.

In the United States, the governance picture is more fragmented. Cybersecurity obligations for the bulk electric system are more mature in some respects through NERC CIP, under the oversight of the Federal Energy Regulatory Commission (FERC) \cite{NERC_CIP_Standards}. Broader risk management guidance is also available through CISA and the NIST Cybersecurity Framework \cite{nist_cybersecurity_framework_2_0}. However, there is no single federal equivalent to the GDPR for general personal data protection. Instead, privacy obligations arise through a patchwork of state-level and sector-specific rules. AI governance remains more reliant on voluntary or soft-law frameworks such as the NIST AI Risk Management Framework \cite{nist_ai_rmf_1_0}. This fragmented structure has practical implications for engineering deployment. A resilience-enhancing AI platform that uses customer-level, health-related, or community vulnerability data may be deployable under one set of conditions in one state and under different or less clear conditions elsewhere. This complicates data integration, benchmarking, and procurement across service territories.

For engineering audiences, the key point is that regulatory context influences technical choices. In the EU, stronger and more unified obligations may increase the burden of compliance, but they also provide a clearer direction toward explainable, traceable, and human-supervised resilience systems. In the US, the greater variation across jurisdictions may allow local experimentation, but it may also hinder standardization and equitable scaling of community-centric resilience tools. Therefore, future resilience research should examine how regulatory structures shape data architecture, human oversight mechanisms, and the long-term deployability of AI-enabled resilience strategies.




\section{Conclusions and Discussions}\label{sec:conclusion}
This paper presents a comprehensive review of community-centric power system resilience, with a focus on AI-enabled methods, critical infrastructure interdependencies, and techno-regulatory considerations. 
The central theme of the paper is disconnect between engineering metrics, social vulnerability indices, and regulatory definitions, leading to a fragmented understanding of resilience as a concept. In addition, there is a tension between data-driven personalization and data privacy, which runs throughout both the AI and legal perspectives on resilience assessment and enhancement methods. The paper also introduces resilience hubs as a key example of a comprehensive, community-centered approach to addressing interdependent and multifaceted vulnerabilities.


Existing resilience metrics are predominantly system-centric (e.g., load served or infrastructure performance) rather than community-centric (e.g., impacts on critical social services and vulnerable populations). Many studies continue to evaluate power system resilience independently of interdependent infrastructures, limiting the ability to quantify cascading impacts. Furthermore, most data-driven approaches focus primarily on outage prediction, while comparatively little attention has been given to modeling community recovery trajectories, service restoration prioritization, and adaptive resilience strategies. Social vulnerability dynamics during outages are also often neglected in resilience assessments.
In addition, current regulatory frameworks largely emphasize reliability compliance rather than measurable community resilience performance, and community-level resilience studies rarely incorporate cyber disruptions affecting grid operations and DER coordination.

 This paper highlights several pressing research gaps in this area. For example: 1) Privacy-preserving federated learning architectures for outage prediction that comply with both GDPR and NERC CIP requirements remain underdeveloped. 2) Formal verification frameworks for ensuring the safety and robustness of DRL-based network reconfiguration agents under cyber-physical attack scenarios, as outlined in NIS2, are still in their early stages of development.
 3) Privacy-preserving resilience analytics remain insufficiently developed, especially for outage prediction and restoration prioritization models. 4) AI-enabled operational methods, especially RL-based approaches for DER coordination, and restoration support, require much stronger safety assurance under cyber-physical attack, telemetry disruption, and extreme out-of-distribution conditions. 5) Resilience hubs and other community-critical facilities need governance-ready design frameworks that integrate lawful data sharing, operational prioritization, and cross-sector accountability. 6) The evaluation of resilience-oriented AI models should move beyond predictive accuracy to include explainability, privacy robustness, and regulatory deployability. 7) Comparative regulatory analysis should be used to inform standardization, so that resilience tools are designed for real-world implementation rather than only for benchmark performance.

Taken together, these issues show that community-centric resilience must be framed as a cyber-physical-social governance challenge. The future research agenda should therefore focus on integrated methods that connect engineering performance with social vulnerability, secure AI system design, and legally compliant operational workflows. Such an agenda would better align resilience investments with the needs of vulnerable communities while improving the trustworthiness and deployability of AI-enabled power system resilience solutions.




\section*{Acknowledgment}
This work is partially supported by the DOE under Award Number DE-EE0010413. Any opinions, findings, conclusions, or recommendations expressed in this material are those of the authors and do not necessarily reflect the views of DOE. The first author acknowledges earlier discussions with Dr. Caisheng Wang on this topic. 
\vspace{-1em}

\begin{singlespace}
\bibliographystyle{IEEEtran}
\bibliography{sources}
\end{singlespace}
\end{document}